\RequirePackage{etoolbox}
\csdef{input@path}{%
	{img/}
}%

\documentclass[12pt]{article}

\usepackage{amsmath}
\usepackage{amssymb}
\usepackage{amsbsy}
\usepackage{amsthm}
\usepackage{mathtools}
\usepackage{mathbbol}
\usepackage{mathrsfs}
\usepackage{bbm} 
\usepackage{graphicx,psfrag,epsf}
\usepackage{enumerate}
\usepackage{natbib}
\usepackage{url} 
\usepackage{float}
\usepackage[skip=5pt]{caption}
\usepackage{subcaption}
\usepackage[normalem]{ulem}
\usepackage{comment}
\usepackage{hhline}
\usepackage{afterpage}
\usepackage{multirow}
\usepackage{adjustbox}
\usepackage{textcomp}
\usepackage{rotating}
\usepackage{tabularx}
\usepackage{xspace}
\usepackage{xr-hyper}
\usepackage[colorlinks,citecolor=blue,urlcolor=blue,filecolor=blue,backref=page]{hyperref}
\usepackage{color,colortbl,soul}
\usepackage[dvipsnames]{xcolor}
\usepackage{blkarray}
\usepackage{tikz}
\usepackage[boxed, vlined]{algorithm2e}
\usepackage[T1]{fontenc}
\usepackage{arydshln}
\usepackage{scalerel,stackengine}
\usepackage{float}
\usepackage{bm}
\usepackage{amsfonts}

\SetAlCapSkip{.5em}

\definecolor{Gray}{gray}{0.9}

\allowdisplaybreaks

\makeatletter
\renewcommand*\env@matrix[1][*\c@MaxMatrixCols c]{%
	\hskip -\arraycolsep
	\let\@ifnextchar\new@ifnextchar
	\array{#1}}
\makeatother

\stackMath
\newcommand\reallywidehat[1]{%
	\savestack{\tmpbox}{\stretchto{%
			\scaleto{%
				\scalerel*[\widthof{\ensuremath{#1}}]{\kern-.6pt\bigwedge\kern-.6pt}%
				{\rule[-\textheight/2]{1ex}{\textheight}}
			}{\textheight}%
		}{0.5ex}}%
	\stackon[1pt]{#1}{\tmpbox}%
}

\makeatletter
\newcommand*{\addFileDependency}[1]{
  \typeout{(#1)}
  \@addtofilelist{#1}
  \IfFileExists{#1}{}{\typeout{No file #1.}}
}
\makeatother

\def\mathcolor#1#{\@mathcolor{#1}}
\def\@mathcolor#1#2#3{%
	\protect\leavevmode
	\begingroup
	\color#1{#2}#3%
	\endgroup
}

\renewcommand{\tilde}[1]{\widetilde{#1}}

\usepackage{algorithmicx}
\usepackage{algpseudocode}

\usepackage{tcolorbox}
\newtcolorbox[auto counter]{reviewcommentinside}[1][]{box align=center,
    width=0.9\textwidth,
    colframe = teal,
    colback=teal!10,
    code={\spacingset{0.9}},
    #1}

\usepackage{tikz}
\usetikzlibrary{bayesnet}

\newcommand{\pkg}[1]{{\normalfont\fontseries{b}\selectfont #1}}
 \let\code=\texttt

\newcommand{\bbE}{\mathbb{E}}

\newcommand{\bX}{\boldsymbol{X}}
\newcommand{\bs}{\boldsymbol{s}}
\newcommand{\bw}{\boldsymbol{w}}
\newcommand{\bc}{\boldsymbol{c}}
\newcommand{\btheta}{\boldsymbol{\theta}}
\newcommand{\wtilde}{\tilde{w}}
\newcommand{\Ytilde}{\tilde{Y}}
\usepackage[margin=1in]{geometry}
\usepackage{mathptmx} 
\usepackage[scaled=.90]{helvet} 
\usepackage{courier} 
\usepackage{graphicx}
\usepackage{amsmath,amssymb}
\usepackage{booktabs}
\usepackage{authblk}
\usepackage{abstract}
\usepackage{setspace}
\usepackage{subcaption}
\usepackage{url}
\usepackage{natbib} 

\title{\textbf{\texttt{svc}: An R package for Spatially Varying Coefficient Models}}
\author[1]{Justice Akuoko-Frimpong}
\author[1]{Edward Shao}
\author[1]{Jonathan Ta}
\affil[1]{Department of Biostatistics, University of Michigan, Ann Arbor}
\date{}

\begin{document}

\maketitle
\noindent\rule{\textwidth}{0.4pt}

\begin{abstract}
Traditional regression models assume stationary relationships between predictors and responses, failing to capture the spatial heterogeneity present in many environmental, epidemiological, and ecological processes. To address this limitation, we develop a scalable Bayesian framework for spatially varying coefficient (SVC) models, implemented in the \pkg{svc} R package (available at \url{https://github.com/jdta95/svc}), 
which allows regression coefficients to vary smoothly over space. Our approach combines three key computational innovations: (1) a subset Gaussian process approximation that reduces the computational burden from $O(n^3)$ to $O(m^3)$ with $m<n$, while maintaining predictive accuracy; (2) a robust adaptive Metropolis (RAM) algorithm that automatically tunes proposal distributions for efficient MCMC sampling of spatial range parameters; 
and (3) optimized linear algebra operations leveraging precomputed distance matrices and Cholesky decompositions to accelerate covariance calculations. We present the model's theoretical foundation, prior specification, and Gibbs sampling algorithm, with a focus on practical implementation for large spatial datasets. Simulation studies demonstrate that our method outperforms existing approaches in computational efficiency while maintaining competitive estimation accuracy. We illustrate its application in an analysis of land surface temperature (LST) data, revealing spatially varying effects of vegetation and emissivity that would be obscured by traditional regression techniques. The \pkg{svc} package provides researchers with a flexible, efficient tool for uncovering and quantifying nonstationary spatial relationships across diverse scientific domains.
\end{abstract}

\section{Introduction}
\label{sec:intro}

Heterogeneity in spatial processes poses challenges for traditional regression models. For example, consider how the relationship between PM2.5 concentrations and factors like traffic density or industrial emissions vary across an urban-rural gradient. Global regression models, assuming constant coefficients, can misrepresent these spatially localized relationships, leading to incorrect conclusions and suboptimal policies.

We propose a Bayesian approach for spatially varying coefficient (SVC) models to address these issues by allowing coefficients to change smoothly over space. Our framework enhances efficiency through three key innovations: a subset Gaussian process (GP) with predictions, a robust adaptive Metropolis (RAM) for optimal random walk sampling, and optimized linear algebra using pre-computed distance matrices and Cholesky decompositions. These improvements make SVC models feasible for large-scale environmental datasets.

%
%
\section{Methods}
\label{sec:methods}
Let $\mathcal{D} \subset \mathbb{R}^2$ denote a spatial domain, $s_i \in \mathcal{D}$ for each $i = 1, \dots, n$ denote a spatial location with collected data, and $\bs = (s_1, \dots, s_n)^\top$ be the vector of all such locations. Then $Y(\bs)$ are univariate dependent variables and $\bX(\bs) = (X_1(\bs), \dots, X_p(\bs))^\top$ are $p \times 1$ vectors of covariates. A linear SVC regression model assumes $Y(\bs)$ are dependent on $\bX(\bs)$ as follows:
\begin{align*}
    Y(\bs) = \sum_{j=1}^p X_r(\bs)w_r(\bs) + \epsilon(\bs),
\end{align*}
where $w_r(\bs)$ are the SVC's corresponding to $X_r(\bs)$ and $\epsilon(\bs)$ are independently and identically distributed multivariate normal measurement errors, i.e. $\epsilon(\bs) \sim \text{MVN}(0, \tau^2 I_n)$. Note that spatially-varying intercept can be included by defining $X_1(\bs)$ as $\mathbf{1}_n$, a vector of length $n$ with every element equal to $1$.

For each $w_r(\bs)$, we assign a GP prior with squared exponential covariance, i.e. $w_r \sim \text{GP}(0, C(\btheta_r))$ where $C(\btheta_r) = \left[C(s_i, s_{i'}; \btheta_r)\right]_{i, i' = 1}^n$. We refer to $C(\cdot)$ as the covariance function, because the covariance of $w_r(s_i)$ and $w_r(s_{i'})$ at locations $s_i$ and $s_{i'}$ is $\sigma_r^2 C(\btheta_r)_{i, i'}$. In particular, we use the squared exponential covariance function where $\btheta_r = (\sigma_r^2, \phi_r)$, $C(\btheta_r) = \sigma_r^2 K(\phi_r)$, and $K(\phi_r) = \left[ K(s_i, s_{i'}; \phi_r) \right]_{i, i' = 1}^n = \left[\exp\{\phi_r^{-1}\|s_i-s_{i'}\|^2\}\right]_{i, i' = 1}^n$,
due to its useful properties such as infinite smoothness and gradual decrease in covariance with distance. The parameter $\sigma_r^2$ is the spatial variance and $\phi_r$ is the spatial range, which indicates how quickly correlation decreases with the squared distance.

We assign inverse-gamma conjugate priors for the variance parameters $\tau^2$ and each $\sigma_r^2$. In addition, each $\phi_r$ is assigned a uniform prior. In summary, $\tau^2 \sim \text{Inv.Gamma}(\alpha_\tau, \beta_\tau)$, $\sigma_r^2 \sim \text{Inv.Gamma}(\alpha_r, \beta_r)$, and $\phi_r \sim \text{Uniform}(l_r, u_r)$.

\subsection{MCMC Algorithm}
\label{sec:mcmc}
The primary function in the \pkg{svc} package, \code{svclm}, samples from the joint posterior distribution of $w_r(\bs), \phi_r, \sigma_r^2,$ and $\tau^2$, for $r = 1, \dots, p$ using a Gibbs sampler. We will explain how each parameter is sampled at iteration $t+1$.

The parameter $\phi_r^{(t+1)}$ is updated via a random walk Metropolis algorithm. Because $\phi_r^{(t)} \in (l_r, u_r)$, each proposal is calculated as 
\begin{align*}
    \phi_r' &= f^{-1}(f(\phi_r^{(t)}) + U) = l_r + \frac{u_r - l_r}{1 + \exp\left\{\log\left(\frac{u_r - l_r}{\phi_r^{(t)} - l_r} - 1\right) + U\right\}},
\end{align*}
where $U$ is sampled from a normal distribution with mean $0$. By default, the proposal standard deviation starts at $1$ and adapts via the RAM algorithm to achieve a target acceptance rate as described by \cite{vihola}.

The sample the SVC's, let $\Ytilde(\bs) = Y(\bs) - \sum_{j\neq r} X_j(\bs)w_j^{(t)}(\bs)$. Then
\begin{align*}
    \frac{\Ytilde}{X_r}(\bs) | w_r^{(t)}(\bs), \tau^2 \sim \text{MVN} \left(w_r^{(t)}(\bs), \tau^{2(t)} \text{diag}\left( \frac{1}{X_r^2(\bs)} \right)\right) := \text{MVN} \left(\mu, \Sigma\right).
\end{align*}
Then assuming the prior $w_r(\bs) \sim \text{MVN}\left(0, C(\btheta_r^{(t)})\right) := \text{MVN}\left(0, \Sigma_0\right)$, we sample $w_r^{(t+1)}(\bs)$ from MVN$(\mu_1, \Sigma_1)$, where $\Sigma_1 = \left(\Sigma_0^{-1} + \Sigma^{-1}\right)$ and $\mu_1 = \Sigma_1 \Sigma^{-1} \frac{\Ytilde}{X_r}(s)$.

We sample $\sigma_r^{2(t+1)}$ from Inv.Gamma$\left(\alpha_r + \frac{n}{2}, \beta_r + w_r^{(t)}(\bs)^{\top} K(\phi_r^{(t)})^{-1} w_r^{(t)}(\bs)\right)$ and $\tau^{2(t+1)}$ from \\ Inv.Gamma$\left(\alpha_t + \frac{n}{2}, \beta_t + \frac{1}{2} \left(Y(\bs) - \sum_{j=0}^p X_r(\bs)w_r^{(t)}(\bs) \right)\right)$.

\subsubsection{Subset Gaussian Process}
\label{sec:subset_GP}

We can perform the above algorithm using the complete data for all $n$ locations to estimate the GP model. The most flop-expensive operation is inverting $C(\btheta_r^{(t)})$, an $(n \times n)$ matrix, for $r = 1, \dots, p$, resulting in an overall cost of $O(pn^3)$ per iteration. We can improve the efficiency of this algorithm with a Guassian process on the subset data with a prediction step as described by \cite{banerjee}.

Consider a set of knots $\bs^* = (s_1^*, \dots, s_m^*)^\top$ which are a subset of locations $\bs$. With the subset GP, we sample $\tau^2$, $\phi_r$, $\sigma_r^2$, and $\bw_r^* = \left[w(\bs_i^*)\right]_{i=1}^m$ which follows a multivariate Gaussian distribution with covariance matrix $C^*(\btheta_r) = \sigma_r^2 K^*(\phi_r) = \left[ \sigma_r^2 \exp\{ \phi_r^{-1} \| s_{i}^* - s_{i}^* \|^2 \} \right]_{i, i' = 1}^m$, for each $r = 1, \dots, p$.

Then for the original set of locations $\bs$, we predict the spatial interpolants $\wtilde_r(\bs) = \bbE[w_r(\bs)|\bw_r^*] = \bc (\bs; \btheta_r) C^{*-1}(\btheta_r)\bw_r^*$, where $\bc (\bs; \btheta_r)$ is an $n \times m$ matrix with $i, j$ element $c_{i, j} = \sigma_r^2 K(s_i, s_j^*; \phi_r) = \sigma_r^2 \exp\{\phi_r^{-1}\|s_i-s_j^*\|^2\}$.

Therefore, using the prediction step, we can replace $w_r(\bs)$ in the original model with $\wtilde_r(\bs)$ for each $r = 1, \dots, p$, but with an overall cost per iteration of $O(pm^3)$, where $m < n$.

\section{Using \pkg{svc}}
\label{sec:using}

To fit a linear SVC regression model using \pkg{svc}, we use the \code{svclm} function. In most cases, we recommend providing the following arguments to \code{svclm}: \code{Y}, \code{X}, \code{coords}, \code{Y\_knots}, \code{X\_knots}, \code{knots}, \code{phi\_lower}, \code{phi\_upper}, and \code{mcmc}. \code{Y} is a vector of length $n$ containing the response variable, \code{X} is the design matrix of size $n \times p$, and \code{coords} is a matrix of size $n \times 2$ containing the spatial coordinates of each location. The arguments \code{Y\_knots}, \code{X\_knots}, and \code{knots} provide the data for the knots that will be used in the subset GP. \code{Y\_knots} is a vector of length $m$ containing the response variable at the knots, \code{X\_knots} is a matrix of size $m \times p$ containing the covariates at the knots, and \code{knots} is a matrix of size $m \times 2$ containing the spatial coordinates of each knot. The arguments \code{phi\_lower} and \code{phi\_upper} are each a vector of length $p$ containing the lower and upper bounds for the uniform distribution priors assigned to spatial range parameters $\phi_r$, respectively. The argument \code{mcmc} is an integer specifying the number of MCMC iterations to run.

The arguments above are sufficient for most cases, but \code{svclm} also has arguments for specifying starting values for $\tau^2$, $\sigma_r^2$, $\phi_r$, and $\bw_r^*$ where $r = 1, \dots, p$. Users can also set the hyperparameters for the inverse gamma priors on $\tau^2$ and $\sigma_r^2$. Lastly, users can set the starting proposal standard deviations for each $\phi_r$ and the target acceptance rate for the RAM.

By default, the starting values for $\tau^2$, $\sigma_r^2$, $\phi_r$, and $\bw_r^*$ are set to $1$, $1$, the midpoint between the corresponding upper and lower bounds, and $0$ for all locations, respectively. The default priors for $\tau^2$ and each $\sigma_r^2$ are all Inv.Gamma$(0.001, 0.001)$ to achieve uninformative priors. The starting proposal standard deviations for each $\phi_r$ are set to $1$ by default, and the target acceptance rate for the RAM algorithm is set to $0.234$ which is asymptotically optimal under most conditions \citep{gelman}.

In all cases, \code{svclm} returns a list containing the following matrices: \code{w\_samples}, an $(\code{mcmc} \times n \times p)$ array of the posterior samples for the SVC's; \code{phi\_samples}, an $(\code{mcmc} \times p)$ matrix of the posterior samples for the spatial range parameters; \code{phi\_acceptance}, an $(\code{mcmc} \times p)$ matrix which is $1$ if the proposal for $\phi_r$ was accepted and $0$ otherwise; \code{sigmasq\_samples}, an $(\code{mcmc} \times p)$ matrix of the posterior samples for the spatial variances; and \code{tausq\_samples}, a $(\code{mcmc} \times 1)$ matrix of the posterior samples for the nugget parameter.

Although knot data can be chosen manually, the \pkg{svc} package also contains a function \code{simpleknots} to automatically select knots for the subset GP. The function takes arguments \code{Y}, \code{X}, \code{coords}, and \code{k}, where \code{k} is an integer and \code{Y}, \code{X}, and \code{coords} match what would be provided to the \code{svclm} function. The function will return a list containing the objects \code{Y\_knots}, \code{X\_knots}, and \code{knots} which can then be passed directly to the \code{svclm} function. The knots are selected by sorting each dimension of coordinates and then selecting every $k$-th coordinate in each dimension.

\section{Simulation}
\label{sec:simulation}

We compared the performance of our package with \pkg{spBayes} using bias, RMSE, and computational time (in seconds), averaged over 1000 simulation replications. In each simulation, we generated data for $441$ locations using the same parameters and obtained knot data using the \code{simpleknots} function with \code{k} set to $2$. Then we used both packages to run an MCMC for $3000$ iterations and predicted the coefficients at all locations using the same knots for both methods. We excluded the first $2000$ iterations of each MCMC chain for burn-in.

Each dataset was generated with $p = 3$, $\sigma_1^2 = \sigma_2^2 = \sigma_3^2 = 1$, $\phi_1 = \phi_2 = \phi_3 = 2$, and $\tau^2 = 0.0001$. The covariates were generated as $X_1 = \mathbf{1}_n$, $X_2 \sim \mathcal{N}(0, 1)$, and $X_3 \sim \mathcal{N}(0, 1)$. True values for the means of $ w_0 $, $ w_1 $, and $ w_2 $ were set to 0, 10, and -5 respectively. The response variable was generated as $Y(\bs) = w_1(\bs) + w_2(\bs) X_2 + w_3(\bs) X_3 + \epsilon(\bs)$.

Table \ref{tab:method-compact} summarizes the results of the simulation. The \pkg{svc} package had better bias and RMSE for $w_1$ but worse bias and RMSE for $w_2$ and $w_3$. Our package also ran more than $3$ times faster on average for the same number of iterations.

\begin{table}[ht!]
\centering
\caption{Comparison of methods based on bias, RMSE, and computation time (seconds).}
\label{tab:method-compact}
\begin{tabular}{|l|cc|cc|cc|c|}
\hline
\textbf{Method} 
& \multicolumn{2}{c|}{$w_1$} 
& \multicolumn{2}{c|}{$w_1$} 
& \multicolumn{2}{c|}{$w_2$} 
& \textbf{Time (s)} \\
\cline{2-7}
& Bias & RMSE 
& Bias & RMSE 
& Bias & RMSE 
& \\
\hline
\pkg{svc} (subset)
& 0.00 & 0.09 
& 1.16 & 1.48 
& -1.30 & 2.08 
& 7.17 \\
\pkg{spBayes} (low rank) 
& -0.01 & 0.21 
& -0.01 & 0.25 
& 0.01 & 0.22 
& 23.26 \\
\hline
\end{tabular}
\end{table}

\begin{figure}[H]
 \centering
 \includegraphics[width=\textwidth]{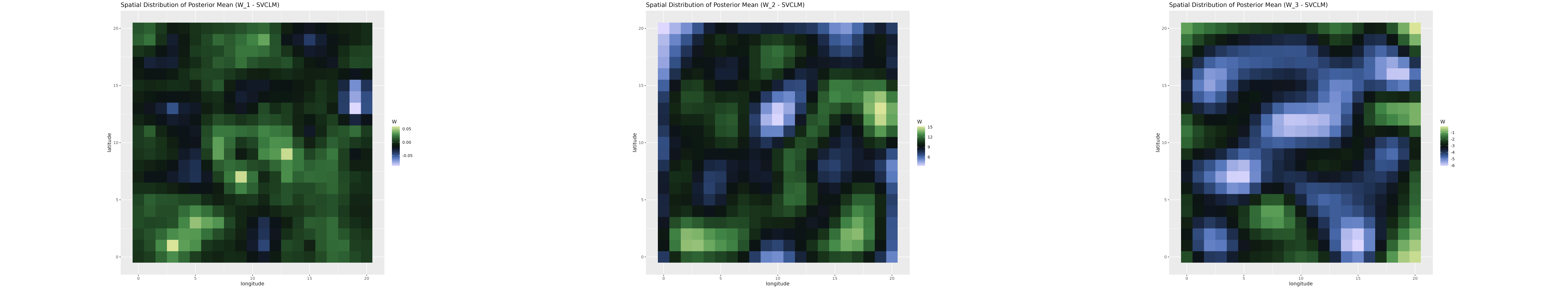}
 \caption{Posterior Mean of SVCLM coefficients for simulation}
 \label{fig:SVCLMA}
 \end{figure}

\section{Data Analysis}
\label{sec:data analysis}
\begin{figure}[H]
 \centering
 \includegraphics[width=6in]{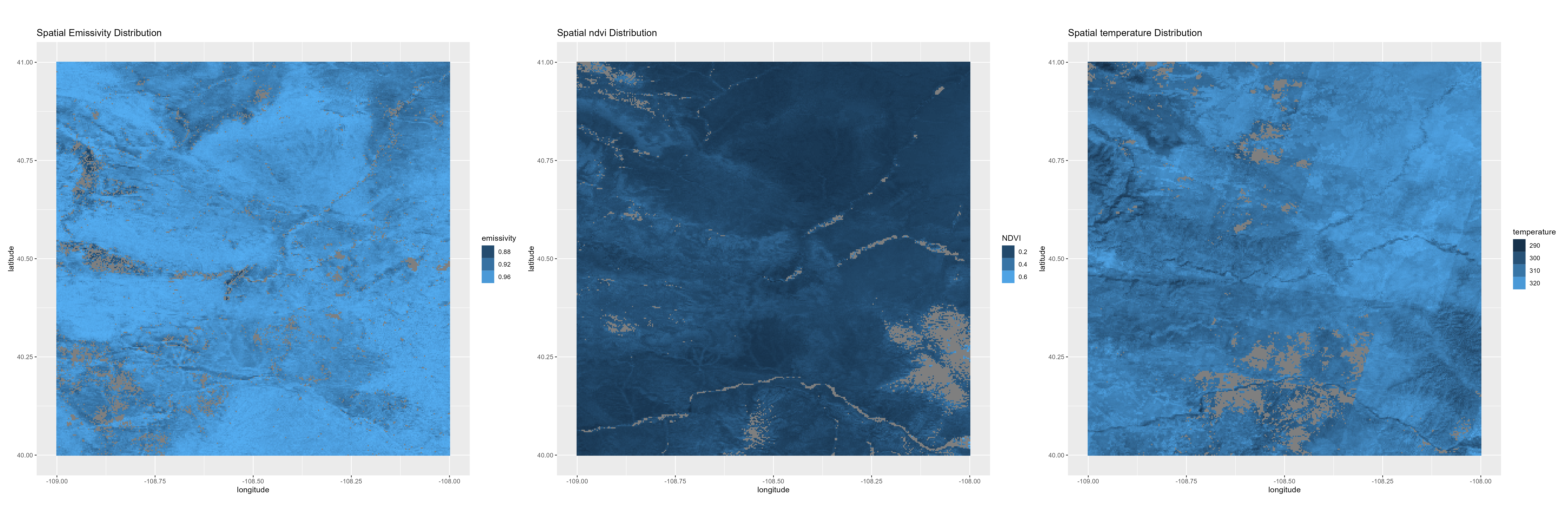}
 \caption{
   Spatial distributions of emissivity, NDVI, and  land surface temperature in the study area. 
 }
 \label{fig:spatialpatterns}
 \end{figure}

We tested \code{svclm} on satellite data with latitude, longitude, temperature, Normalized Difference Vegetation Index (NDVI); which measures greenness, and emissivity which represents a surface's efficiency in emitting thermal radiation \citep{hulley_new_2008, hulley_aster_2008, hulley_north_2009, hulley_validation_2009, hulley_generating_2011, hulley_quantifying_2012, nasa_jpl_aster_2014, hulley_span_2015}.
The data had 111,556 observations.
We used LST as a univariate outcome and used emissivity and NDVI as covariates.
Figure \ref{fig:spatialpatterns} below reveals the spatial  distributions of the variables in the dataset.
Emissivity ranged from 0.88 to 0.96, NDVI ranged from 0.2 to 0.6  and LST, from 290 to 320.  
 Cooler regions often align with greener (higher NDVI) areas — vegetation can reduce surface temperature through evapotranspiration. 
\subsection{Model}
We modeled (LST) using an SVC framework:
\begin{align*}
\text{Temp}(\bs) &= w_1(\bs) + w_2(\bs)\cdot\text{NDVI} + w_3(\bs)\cdot\text{Emissivity} + \epsilon(\bs),
\end{align*}
with priors as specified in Section \ref{sec:methods}.

We specified weakly informative priors: $\tau^2 \sim \text{Inv-Gamma}(0.001, 0.001)$, \newline $\sigma_r^2 \sim \text{Inv-Gamma}(0.001, 0.001)$, and $\phi_r \sim \text{Uniform}(0.001, 500)$, for each $r = 1, \dots 3$.

The computational implementation featured subset GP. We selected 1 knot per 100 grid cells ($k=10$) using the  $\texttt{simpleknots()}$ function. Removed knots with missing data.
\subsection{Results}
%
%
Below is a plot of the posterior means of the coefficients at each location. The means were computed after burn-in.
\begin{figure}[H]
 \centering
 \includegraphics[width=\textwidth]{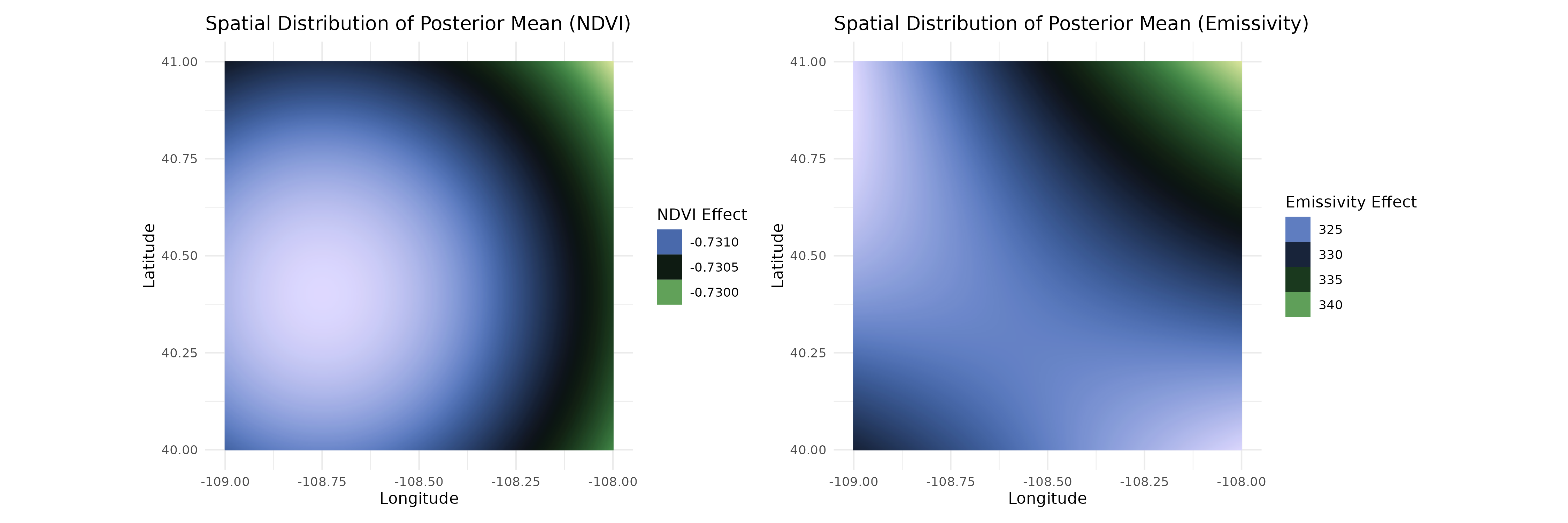}
 \caption{Spatially varying coefficient surfaces for (A) NDVI and (B) Emissivity effects on land surface temperature}
 \label{fig:posterior means}
 \end{figure}
 The posterior mean NDVI coefficients show strong negative effects (cooling) across the entire study region, with values ranging from -0.7310 to -0.7300. 
 Each unit increase in NDVI corresponds to about $0.73^\circ C$ decrease in land surface temperature across space.
 The emissivity coefficients ranged from 325 to 340. These values revealed strong positive associations  which implies that across space, a unit increase in emissivity increases temperature.
 
 We generated predicted temperature values across the entire spatial domain using the posterior mean estimates of the SVC's. 
 These predictions were computed for all locations with complete predictor data, regardless of whether the actual temperature was observed. 
 The predicted values were obtained by multiplying the mean coefficient estimates with the corresponding covariate values at each location. 
 We then visualized both the observed and predicted temperature surfaces side by side to assess model performance and spatial prediction accuracy.

 \begin{figure}[H]
 \centering
 \includegraphics[width=\textwidth]{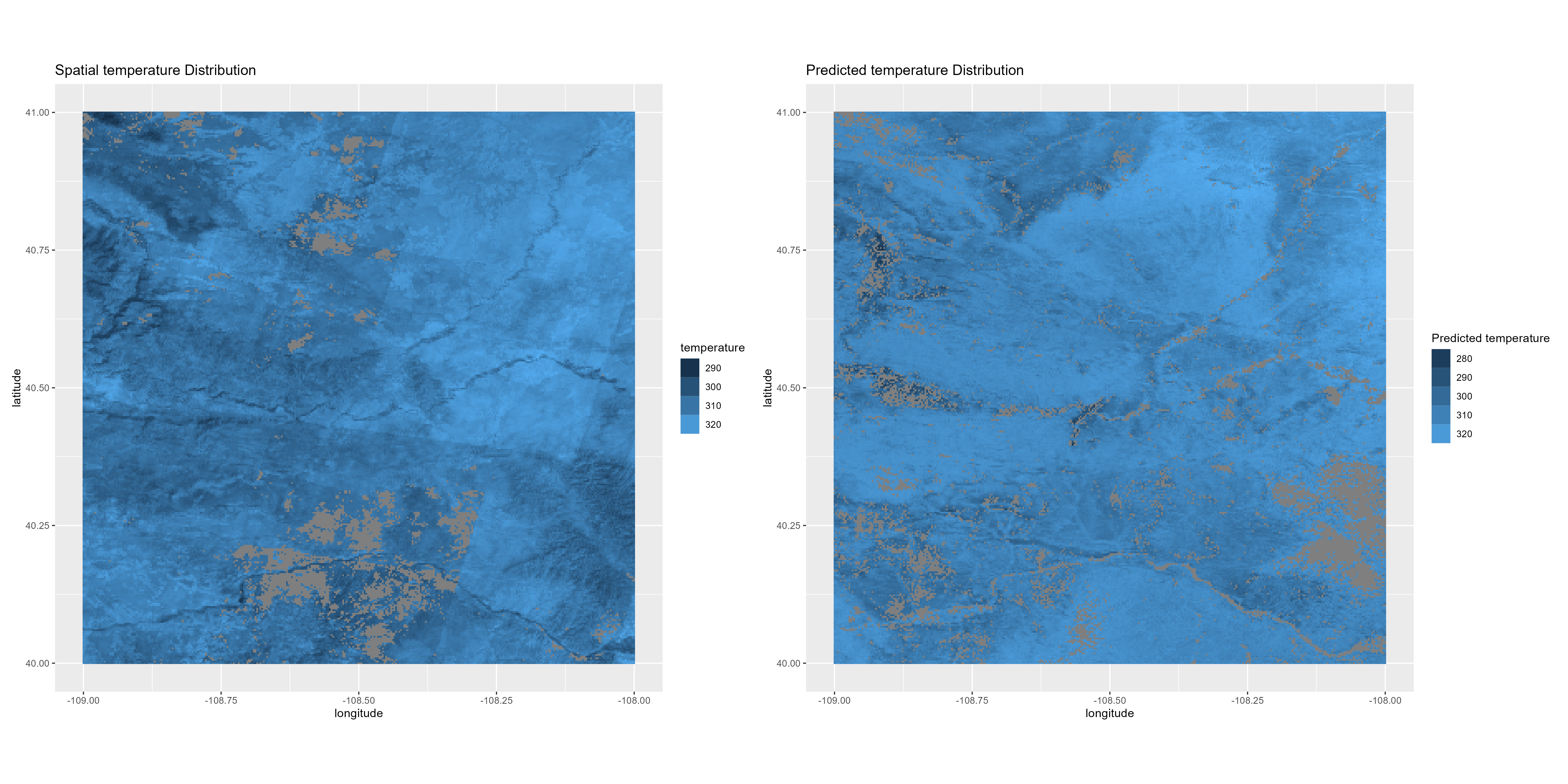}
 \caption{Side by side plot of temperature and their predicted values across space. Gray indicates missing values.}
 \label{fig:predicted temperature}
 \end{figure}
 
 Note that the gray spaces in the plot are the locations with missing data. We can predict LST at locations with missing LST but cannot predict at locations with missing covariates; NDVI and emissivity.The figure comparison reveals the model’s ability to capture spatial patterns in LST, even in regions where observations were missing but predictor data were available.
 The range of values between the actual and predicted were close.
 
\section{Conclusion}
\label{sec:Conclusion}
The \pkg{svc} package provides a flexible and efficient framework for fitting SVC models that capture heterogeneous spatial processes. By leveraging the subset GP and predictions, RAM adaptation, and various sampling optimizations, we have made SVC models more computationally feasible for large-scale spatial datasets.

Compared to \pkg{spBayes}, our method performs more efficiently and has lower bias and RMSE for the spatial random intercept, but has higher bias and RMSE for all other SVC's. However, \pkg{svc} did run much faster than \pkg{spBayes} for the same number of iterations.

One limitation of our method is that it requires complete data at knot locations, whereas \pkg{spBayes} does not require any data at selected knot locations.

Further improvements to the package could include allowing for different covariance functions and implementing parallelization for sampling $\phi_r$ and $\sigma_r^2$ for $r = 1, \dots, p$ since the sampling can be done independently for each $r$.

\newpage

\bibliographystyle{agsm}

\bibliography{bibliography}
\end{document}